\documentclass[12pt,preprint,a4paper]{aastex}
\usepackage{times}
\usepackage{graphics,epsf}
\usepackage{amsmath}                % American Mathematical Society package
\usepackage{amsfonts}               % American Mathematical Society fonts
\usepackage{amssymb}                % American Mathematical Society symbol
\usepackage{epsfig}                 % EPS figures
\usepackage{graphicx}
\usepackage{url}
\usepackage{tabularx}
\usepackage{booktabs}
\usepackage{hyperref}
\usepackage[dvipsnames]{xcolor}
\def \s{~\rm{s}}

\def \km{~\rm{km}}

\def \yr{~\rm{yr}}

\def \kpc{~\rm{kpc}}

\begin{document}

\title{A jet-driven dynamo (JEDD) from jets-inflated bubbles in cooling flows}

\author{Noam Soker}

\affil{Department of Physics, Technion -- Israel Institute of
Technology, Haifa 32000, Israel; soker@physics.technion.ac.il}
%\author{Avishai Gilkis\altaffilmark{1} and Noam Soker\altaffilmark{1}}
%\altaffiltext{1}{Department of Physics, Technion -- Israel
%Institute of Technology, Haifa 32000, Israel;
%agilkis@tx.technion.ac.il; soker@physics.technion.ac.il}

\begin{abstract}
I suggest that the main process that amplifies magnetic fields in cooling flows in clusters and group of galaxies is a jet-driven dynamo (JEDD). The main processes that are behind the JEDD is the turbulence that is formed by the many vortices formed in the inflation processes of bubbles, and the large scale shear formed by the propagating jet. It is sufficient that a strong turbulence exits in the vicinity of the jets and bubbles, just where the shear is large.    The typical amplification time of magnetic fields by the JEDD near the jets and bubbles is approximately hundred million years. The amplification time in the entire cooling flow region is somewhat longer.
The vortices that create the turbulence are those that also transfer energy from the jets to the intra-cluster medium, by mixing shocked jet gas with the intra-cluster medium gas, and by exciting sound waves.
The JEDD model adds magnetic fields to the cyclical behavior of energy and mass in the jet-feedback mechanism (JFM) in cooling flows.
\end{abstract}


% ==========================================================
\section{INTRODUCTION}
\label{sec:introduction}
% ==========================================================

The advancement in observations and in numerical simulations in recent years have brought to the recognition that jet-inflated bubbles play the dominant role in heating the intracluster medium (ICM) in cooling flows in clusters and groups of galaxies.
Only five years ago several processes were discussed as candidates for heating the ICM (e.g., see reviews by
\citealt{Fabian2012} and \citealt{McNamaraNulsen2012}). Although it was agreed by most researchers that the
heating must operate through a negative feedback mechanism (e.g., \citealt{Farage2012, FujitaOhira2013, Gasparietal2013, Pfrommer2013}),
there were several alternative heating processes. Recent results from the Hitomi X-ray telescope, for example, suggest that turbulence
is unlikely to heat the ICM in the Perseus cluster \citep{Hitomi2016}, a result anticipated by three-dimensional (3D) hydrodynamical
simulations of jet-inflated bubbles \citep{HillelSoker2016, HillelSoker2017}.
Energetically it is possible for the observed turbulence to heat the ICM in perseus, as discussed by
\cite{Hitomi2016} in relation to the study by \cite{Zhuravlevaetal2014}, but \cite{Hitomi2016} view this as unlikely.
My view is that these new observations and simulations leave us with the jet-feedback mechanism (JFM), where jets inflate
bubbles and heat the ICM (see review by \citealt{Soker2016}).

Different processes have been proposed to channel energy from the jets and the bubbles they inflate to heat the ICM (e.g., \citealt{Gasparietal2013, Lietal2017}), including cosmic rays and thermal conduction (e.g., \citealt{GuoOh2008}), sound waves \citep{Fabian2012, Fabianetal2017} that are excited by the inflation process of bubbles \citep{SternbergSoker2009}, and mixing of hot post-shock gas from the bubble with the ICM (e.g., \citealt{BruggenKaiser2002, Bruggenetal2009, GilkisSoker2012, HillelSoker2014, BanerjeeSharma2014, Prasadetal2015, HillelSoker2016, HillelSoker2017, Gasparietal2017}).

As the jets propagate through the ambient medium and inflate bubbles, they form many vortices inside the bubbles and in the surroundings ambient gas. These vortices mix the hot gas inside the bubbles with the ICM, and heat the ICM, in what is termed mixing-heating (e.g., \citealt{GilkisSoker2012, HillelSoker2016, YangReynolds2016b, Gasparietal2017}). A turbulent region is developed around the propagating jets.

In the present study I examine the possibility that the turbulent regions that are formed around the propagating jets and bubbles amplify magnetic fields, i.e., form a dynamo.
Observations (see review by \citealt{CarilliTaylor2002}), such as faraday rotation measures, indicate the presence of relatively strong magnetic fields in cooling flow clusters (e.g., \citealt{GeOwen1994, Tayloretal2002}).
There are many papers on the evolution of the magnetic fields in cooling flow clusters
(e.g., \citealt{CarilliTaylor2002, Soker2010}). Some consider the amplification of magnetic fields by the inflow (e.g., \citealt{SokerSarazin1990, DuboisTeyssier2008}),
in some cases aided by angular momentum of the ICM \citep{Godonetal1998}. The magnetic fields suppress heat conduction in the ICM (e.g., \citealt{BregmanDavid1988}),
and can play a role in the evolution of optical filaments (e.g., \citealt{Godonetal1994, Fabianetal2016}), as well as in X-ray filaments \citep{Sarazinetal1992}.

The better understanding of the heating mechanism, and in particular the JFM, motivates me to reconsider
the amplification of ICM magnetic fields by the jets and the bubbles they inflate. There are several earlier studies of this kind.
\cite{Bicknelletal1990} performed 3D magnetohydrodynamic simulations and find the mixing of gas and magnetic fields from the jet with
the ambient gas. They attribute the mixing to Kelvin-Helmholtz instabilities. \cite{Koideetal1996} performed 3D magnetohydrodynamic
simulations to study the bending of jets by an ambient magnetic field.

I note that \cite{GoldshmidtRephaeli1993} concluded that turbulent dynamo models cannot account for the observed ICM magnetic fields.
In the present study the turbulent dynamo growth rate is enhanced by the presence of a shear formed by the propagating jets.
Amplification of magnetic fields by the shear of propagating jets was discussed by \cite{Soker1997}.

\cite{Gaibleretal2009} performed 3D magnetohydrodynamic simulations of jets propagating through an ambient gas, and examined the magnetic field structure. They found that the flow near the jet head converts the poloidal magnetic field into toroidal magnetic field. There is a toroidal shear near the jet head, i.e., perpendicular to the propagation direction of the jet. They commented that the amplification of magnetic fields is related to the solar dynamo, where in both cases there is a toroidal sher, and the solar convection is replaced by jet-driven cocoon turbulence. In the present study the shear is along the propagation direction of the jet.

\cite{Duboisetal2009} found in their 3D magnetohydrodynamic simulations that the jet-induced turbulence strongly amplifies the magnetic fields in the ICM. In their simulations this takes place mainly when the jets turn the cluster to non-cooling flow one. Here I am interested in the role of jets in cooling flow clusters.

Then Huarte-Espinosa and collaborators conducted 3D magnetohydrodynamic simulations (\citealt{HuarteEspinosa2011a, HuarteEspinosa2011b, HuarteEspinosa2012}) and demonstrated how both turbulence that is formed by the propagating jet (also \citealt{YangReynolds2016a}), and the shear along the propagation direction of the jet, amplify magnetic fields in the ambient gas and in the cocoon (the post-shock jet's material). They basically obtained a dynamo based on 3D turbulence and shear along the propagation direction of the jet.

In this study the earlier results of dynamo excited by propagating jets (\citealt{Duboisetal2009, Gaibleretal2009, HuarteEspinosa2011a, HuarteEspinosa2011b}), the results of the heating-mixing simulations (e.g., \citealt{HillelSoker2016, HillelSoker2017}), and the new study by \cite{Singhetal2017} of the enhancement of small-scale turbulent dynamo by large-scale shear, are put together into a coherent picture of what I term a jet-driven dynamo (JEDD) that takes place as part of the JFM in cooling flows, and possibly in other environments where the JFM occurs.

% ==========================================================
\section{THE JET-DRIVEN DYNAMO (JEDD)}
\label{sec:bubble}
% ==========================================================

The derivation presented here is based on the recent analytical and numerical study of \cite{Singhetal2017}, where more details and earlier references can be found.
The case presented here is that of a turbulence produced by the shear.
We work in cylindrical coordinate system. The $z$ axis is taken along the momentarily symmetry axis of the jet, and the coordinate $r$ is perpendicular to $z$.

There is a shear as a result of a velocity gradient $S_L=- d v_z/dr >0$, formed by the jet that is propagating along the $+z$ direction. The shear can be approximated as
\begin{equation}
S_L \simeq \frac{v_{\rm h,jet}}{R_b} = \frac{1}{1.4 \times 10^7}
 \left( \frac{v_{\rm h,jet}}{700 \km \s^{-1}} \right)
 \left( \frac{R_b}{10 \kpc} \right)^{-1} \yr ^{-1} ,
\label{eq:shear1}
\end{equation}
where $v_{\rm h,jet}$ is the propagation velocity of the head of the jet, and $R_b$ is the radius of the bubble inflated by the jet.
Here and in what follows we will scale quantities according to the simulations of jet-inflated bubbles conducted by \cite{HillelSoker2016} in three dimensions, and by \cite{GilkisSoker2012} and \cite{HillelSoker2014} in two dimensions.
The small-scale shear rate is defined as
\begin{equation}
s_t = \frac{u_{\rm rms}}{l_f} =
 \frac{1}{1.0 \times 10^7}
 \left( \frac{u_{\rm rms}}{500 \km \s^{-1}} \right)
 \left( \frac{l_f}{5 \kpc} \right)^{-1} \yr ^{-1} ,
\label{eq:shear2}
\end{equation}
where $l_f$ is the scale of the turbulence, $u_{\rm rms}=\sqrt{\overline {u^2}}$, and $u$ are the fluctuations of the fluid velocity. The scaling is from figure 1 of \cite{HillelSoker2016}.
In the case of turbulence produced by shear, the small-scale shear rate $s_t$ cannot be
much smaller than the large-scale shear. Comparing equations (\ref{eq:shear1}) and (\ref{eq:shear2}) we see that this condition is met in the case of turbulence produced by jet-inflated bubbles.

\cite{Singhetal2017} confirmed the finding of \cite{Kolokolovetal2011} that the linear shear increases the small-scale dynamo growth rate. In their simulation \cite{Singhetal2017} find that the fluctuating magnetic field growth rate increase as $\gamma \propto \vert S \vert$.
For $S_L=0.009 c_s k_f$, where $c_s$ is the sound speed and $k_f = 2 \pi/ \lambda_f$ is wave number of the velocity fluctuations of the turbulence, \cite{Singhetal2017} find that the magnetic field reaches equipartition (where the magnetic pressure is about equal to the turbulent pressure) at a time of $\tau_{\rm eq} \simeq 190 (k_f u_{\rm rms})^{-1}$.
For $\lambda_f \approx 5 \kpc$ and $c_s \approx 800 \km \s^{-1}$, the shear in the present study (according to equation \ref{eq:shear1}) is $S_L= 0.07 c_s k_f$. The growth rate in the present study can be several time larger. I crudely estimate the time to reach equipartition under the assumptions made by \cite{Singhetal2017}, by taking $\tau_{\rm eq} \propto \gamma^{-1} \propto S^{-1}_L$. This gives
\begin{equation}
\tau_{\rm eq} \approx 190 (k_f u_{\rm rms})^{-1}
   \left( \frac{S_L}{0.009 c_s k_f} \right)^{-1}
   \simeq 3 S^{-1}_L
    \left( \frac{c_s}{2u_{\rm rms}} \right) .
\label{eq:taueq}
\end{equation}

Equation (\ref{eq:taueq}) is a very crude estimate. For the operation of the dynamo the turbulence should be a three-dimensional one. Namely, the magnetic fields cannot be amplified in meridional planes. Therefore, there must be a deviation from pure axisymmetry. In most cases, bubble pairs in cooling flows possess departure from axisymmetry, such as the two bubbles and the center are not along a straight line (e.g., A2597, \citealt{Clarkeetal2005};  HCG~62 \citealt{Gittietal2010}), the two consecutive bubble pairs do not share the same symmetry axis (e.g., NGC~5813, \citealt{Randalletal2015}; Hydra~A, \citealt{Wiseetal2007}), or the jets precess (e.g., MS~0735+7421, \citealt{Gittietal2007}).
Such departure from axisymmetry ensures 3-dimensional turbulence. However, the efficiency of the dynamo might be reduced.

{{{{ One requirement for the proposed dynamo operation is the excitation of a turbulence by the propagating jet in the vicinity of the jet and the bubble(s) it inflates. Namely, the turbulence is required where the shear due to the outward propagating jet is large, along the jet and its vicinities. As evident from different simulations that were mentioned in section 1, jets excite turbulence in the required region. In that respect, the finding that turbulence cannot heat the ICM in the Perseus cluster \citep{Hitomi2016} is not a problem, as that result was anticipated by simulations of jet-inflated bubbles that show efficient excitation of turbulence in the vicinity of jets and the bubbles they inflate \citep{HillelSoker2017}. Neither the finding of \cite{YangReynolds2016a} that AGN-driven turbulence affects only regions directly influenced by the jets is a problem. As evident from figure 4 of \cite{YangReynolds2016a}, the turbulence is excited where it is required for the JEDD. }}}}

I summarize by presenting a crude estimate of the timescale that the jet-driven dynamo (JEDD) brings the magnetic field to equipartition, assuming they did not start from a too low value. Namely, the time the dynamo amplifies the magnetic filed in the vicinity of the jets and bubbles in cooling flow clusters by a few orders of magnitude is crudely given by
\begin{equation}
\tau_{\rm eq,CF} \approx 10 \frac{R_b}{v_{\rm h,jet}} \simeq 10^8
 \left( \frac{R_b}{10 \kpc} \right)^{-1}
 \left( \frac{v_{\rm h,jet}}{700 \km \s^{-1}} \right)^{-1} \yr .
\label{eq:taueqcf}
\end{equation}
The amplification time in the entire cooling flow region is somewhat longer, and will have to be determine by 3D numerical simulations that cover the entire cooling flow region and run for several jet activity periods.

% ==========================================================
\section{SUMMARY}
\label{sec:summary}
% ==========================================================

From simple arguments I concluded in section \ref{sec:bubble} that the interaction of jets and the bubbles they inflate in the ICM and ISM drives a dynamo
that is based on turbulence and shear {{{{ in the vicinity of the jets and bubbles. }}}}
Numerical simulations over the years (see section \ref{sec:introduction}) demonstrated the existence of shear along the propagation direction of the jet and the formation of turbulent regions around the jet and the bubble it inflates. The simple arguments I used are based on the recent paper by \cite{Singhetal2017}. The amplification of the magnetic fields by such a dynamo was obtained for specific cases in 3D hydrodynamical simulations \citep{Duboisetal2009, HuarteEspinosa2011a, HuarteEspinosa2011b}. Here I derived a general expression (eq. \ref{eq:taueqcf}) for the amplification time of magnetic fields in the ICM, based on more recent simulations of the heating of the ICM with jet-inflated bubbles. Note that equation (\ref{eq:taueqcf}) gives the amplification time near the jets and bubbles, and that the average amplification time of the magnetic fields in the entire cooling flow region is somewhat longer.
Nonetheless, I argue that the jet-driven dynamo (JEDD) is the main process of amplifying magnetic fields in cooling flows.

{{{{ The JEDD has some limits, and its effectiveness might not be as large as expressed in equation (\ref{eq:taueqcf}). First, there is the question whether the 3D turbulence in the vicinity of the jets and bubbles scan amplify the magnetic fields on a time scale of $\approx 10^8 \yr$. The simulations performed by Huarte-Espinosa and collaborators (\citealt{HuarteEspinosa2011a, HuarteEspinosa2011b}), that show that jet-induced turbulence and the shear along the jet amplify magnetic fields, suggest that the JEDD can indeed be efficient in the vicinity of the jets and bubbles. The second challenge the JEDD should overcome is the distribution of the strong magnetic fields that were amplified along the propagation directions of the jets in the entire ICM. This can be best achieved if the axis of the jets that are launched by the central active galactic nucleus changes over time. As discussed above, there are indications that this is indeed the case. The next step would be to conduct 3D magnetohydrodynamic simulations that incorporate the detail physics of the turbulence and shear along the propagating jets, and at the same time follow the evolution of a large volume of the ICM over several feedback cycles. These are very computationally demanding calculations. }}}}

The main outcome of my claimed JEDD is that magnetic fields are incorporated to the cyclical behavior of the JFM in cooling flows, together with the energy and mass.

The energy cycle is the most important one, as it sets the feedback. The jets are energized by the accretion of mass onto the central super-massive black hole, and in turn the jets inflate bubbles and heat the ICM. Many vortices are formed inside and outside the bubbles during the inflation of bubbles. These vortices excite sound waves and mix hot gas from the bubbles with the ICM; both processes transfer energy from the propagating jet to the ICM (section \ref{sec:introduction}). As well, the vortices form turbulent regions around the propagating jet that together with the shear along the propagation direction amplify the magnetic fields in the JEDD.

Mass is an important component of the feedback cycle (for a review see \citealt{Soker2016}). Non negligible mass in cold clumps that are formed in the ICM flows inward in a rain (precipitation;  chaotic cold accretion) of cold clouds to feed the central active galactic nucleus (AGN)
(e.g., \citealt{PizzolatoSoker2005, Gaspari2012, Gasparietal2015, Prasadetal2015, VoitDonahue2015, ChoudhurySharma2016, YangReynolds2016b, Voitetal2017, Prasadetal2017, Gasparietal2017}), in what is termed the cold feedback mechanism. In the cold feedback mechanism the perturbations that the jets excite (e.g., \citealt{PizzolatoSoker2005}), such as by lifting cool, low-entropy gas in the wakes of  buoyantly-rising bubbles \citep{McNamaraetal2016} or forming the turbulence (e.g., \citealt{Gasparietal2017}), serve as the seeds for the nonlinear condensations from which the cold filament and clumps form.
In turn, observations show that AGN can launch jets that are wide and carry large amount of mass (e.g., \citealt{Moe2009, Aravetal2013, ChamberlainArav2015, Cheungetal2016, Harrisonetal2014, Williamsetal2016}).

The present study adds the magnetic field to this cycle. Magnetic fields are lost via reconnection and Ohmic dissipation, and hence must be replenished. In that sense the magnetic cycle is more  similar to the energy cycle than to the mass cycle. Mass is basically conserved in the cycle. Only small amounts of mass are lost to star formation and to the AGN. Mass then flows from the outer parts of the ICM to fill-in. Energy is lost by radiative cooling, mainly in the X-ray band, and is replenished by the accretion process on to the AGN. Magnetic fields must also be replenished. Here I suggest that the main process that amplifies magnetic fields in cooling flows is the JEDD. The main process that is behind the JEDD is the turbulence that is formed by the many vortices formed in the inflation processes of bubbles. The same vortices also mix the hot shocked jet gas with the ICM \citep{HillelSoker2016, Gasparietal2017} and excite sound waves in the ICM \citep{SternbergSoker2009}, both of which heat the ICM (see section \ref{sec:introduction}).

% %%%%%%%%%%%%  Refrences %%%%%%%%%%%%%%%%%%%%%%%%%%%%%%%%%%%%%%%%%%%%%%%%%%%%%%%%%%%%%%%%%%%%%%%%%%%%%%%%%%%%%

\label{lastpage}

\end{document}